\documentclass[a4paper,12pt]{article}
\pdfoutput=1 

\usepackage{jheppub} 

\usepackage[T1]{fontenc} 
\usepackage{xcolor}
\usepackage{physics}
\usepackage{multirow}
\usepackage{subcaption}
\usepackage{slashed}
\usepackage{esint}
\usepackage{color}
\usepackage{ulem}

\graphicspath{ {graphics/} }

\makeatletter
\gdef\@fpheader{}
\makeatother

\title{\boldmath Searching Dark Photons using displaced vertices at Belle II -- with backgrounds}

\author[1]{Joerg Jaeckel}
\author[1,2,3]{and Anh Vu Phan}
\affiliation[1]{Institute for Theoretical Physics, Heidelberg University, 69120 Heidelberg, Germany}
\affiliation[2]{Institute for Mathematics, Astrophysics and Particle Physics, Radboud University, 6500 GL Nijmegen, The Netherlands}
\affiliation[3]{Nikhef, Science Park 105, 1098 XG Amsterdam, The Netherlands}
\emailAdd{jjaeckel@thphys.uni-heidelberg.de}
\emailAdd{anhvu.phan@ru.nl}

\abstract{
Dark photons in the MeV to GeV range with kinetic mixing of the order of $\lesssim 10^{-4}-10^{-3}$ can be produced in significant numbers at low energy colliders such as Belle II. Their decay length can be macroscopic raising the hope for a fairly clean search via displaced vertices as proposed in~\cite{Ferber:2022ewf}. However, even this is not background free. Here, we calculate and discuss problematic backgrounds from displaced photon conversion and discuss their potential impact on the sensitivity. 
In addition we also briefly consider the dangers of prompt backgrounds.
}

\begin{document} 
\newcommand{\eV}{\, \text{eV}}
\newcommand{\MeV}{\, \text{MeV}}
\newcommand{\GeV}{\, \text{GeV}}
\newcommand{\fm}{\, \text{fm}}
\newcommand{\cm}{\, \text{cm}}

\newcommand{\mA}{\mathcal{A}}
\newcommand{\mH}{\mathcal{H}}
\newcommand{\mI}{\mathcal{I}}
\newcommand{\mL}{\mathcal{L}}
\newcommand{\mM}{\mathcal{M}}
\newcommand{\mO}{\mathcal{O}}
\newcommand{\mV}{\mathcal{V}}

\newcommand{\q}{\ ,\quad}
\newcommand{\e}[1]{\cdot 10^{#1}}
\newcommand{\nn}{\notag\\}
\newcommand{\figref}[1]{figure~\ref{#1}}
\renewcommand{\eqref}[1]{Eq.~(\ref{#1})}

\newcommand{\rmin}{r_{\text{min}}}
\newcommand{\rmax}{r_{\text{max}}}

\newcommand{\twocases}[2]{
\begin{cases}
	#1\\
	#2
\end{cases}}
\newcommand{\threecases}[3]{
	\begin{cases}
		#1\\
		#2\\
		#3
\end{cases}}
\newcommand{\fourcases}[4]{
	\begin{cases}
		#1\\
		#2\\
		#3\\
		#4
\end{cases}}
\maketitle
\flushbottom

\section{Introduction} 
\label{sec:intro}
The search for feebly interacting particles (FIPs) is one of the most active frontiers in particle physics, cf., e.g.~\cite{Beacham:2019nyx,Agrawal:2021dbo,Antel:2023hkf}.
Their feeble interactions require experiments to ideally feature very high luminosities and good control over backgrounds. Luckily FIPs often also feature signals that are distinct from those expected in the Standard Model. One of those is the existence of displaced vertices. Consequently, displaced vertices have been exploited in different forms for novel search strategies at colliders, cf. e.g.~\cite{Batell:2009yf,Essig:2009nc,Morrissey:2014yma,Chou:2016lxi,Feng:2017uoz,Gligorov:2018vkc,FASER:2018bac,Alimena:2019zri,Bauer:2019vqk,Aielli:2019ivi,Alpigiani:2020iam,Curtin:2018mvb,Filimonova:2019tuy,LHCb:2019vmc,CMS:2019buh,MammenAbraham:2020hex,Dreyer:2021aqd,Schafer:2022shi,Ferber:2022ewf,Bandyopadhyay:2022klg,Bandyopadhyay:2023lvo,Rygaard:2023dlx} but also in fixed target experiments and beam dumps, cf. e.g.~\cite{CHARM:1985anb,Konaka:1986cb,Riordan:1987aw,Bjorken:1988as,Davier:1989wz,Blumlein:1990ay,Blumlein:1991xh,Bjorken:2009mm,Batell:2009di,Gninenko:2011uv,Andreas:2012mt,Bonivento:2013jag,Dobrich:2015jyk,Alekhin:2015byh,NA64:2018lsq,Baldini:2021hfw,Tastet:2020tzh,NA62:2019meo}. 

Still, new ideas to utilize displaced vertices continue to be developed and implemented, complemented by an evolving experimental landscape.  A particularly interesting one~\cite{Ferber:2022ewf} is to take advantage of the currently running Belle II experiment. There, displaced vertex searches for dark photons seem very promising~\cite{Ferber:2022ewf} (see also~\cite{Bandyopadhyay:2022klg,Bandyopadhyay:2023lvo} for the case of partially invisibly decaying and other more general~\cite{Bauer:2018onh} dark photons). They may provide sensitivity~\cite{Ferber:2022ewf} in a large range of interesting parameter space that currently is untested because the lifetime would be too short for beam dumps and the rate is too small to see a significant signal above background in searches where a displaced vertex is not resolved. 

Yet, even displaced vertices are not completely background free and some care needs to be taken. In the case of dark photons at Belle II, an important background is given by photons produced in QED processes that are then converted into electron positron pairs inside the detector material (thereby being produced ``displaced'').
Another, in principle reducible, background is from prompt electron-positron pair production, where the vertex is mis-reconstructed. 
In the present paper we want to build on the study of~\cite{Ferber:2022ewf} by a more careful calculation of these backgrounds and a discussion of suitable cuts that can be used to reduce them. We also briefly discuss the required suppression of the reducible prompt background. We then use these results for an updated sensitivity estimate to dark photons.

In Sec.~\ref{sec:modelandsignal} we briefly recall the model used and discuss the expected signal. The main calculation of the photon conversion background is then performed in Sec.~\ref{sec:background}. There, we also obtain the rate of prompt electron positron production. Additional details of our modelling for the background calculation are given in Appendix~\ref{app:conversion}. The resulting impact on the sensitivity is then discussed in Sec.~\ref{sec:results}. Sec.~\ref{sec:conclusion} provides a brief summary and conclusions.

\section{The dark photon model}
\label{sec:modelandsignal}

\subsection{Dark photons and their signal}
In this paper we are discussing a search strategy for dark photons, i.e. massive U(1) vector bosons, interacting with the Standard Model through mixing with the Standard Model U(1)~\cite{Okun:1982xi,Foot:1991kb,Holdom:1985ag} (cf. also, e.g.~\cite{Jaeckel:2010ni,Jaeckel:2012mjv,Fabbrichesi:2020wbt,Agrawal:2021dbo,Antel:2023hkf} for reviews of this and other portals). To be explicit we use the following Largrangian,
\begin{eqnarray}
    {\mathcal{L}}&=&{\mathcal {L}}_{\rm SM}+ {\mathcal{L}}_{\rm DP}
    \\\nonumber 
    &=&{\mathcal{L}}_{\rm SM}+ \frac{1}{4}(X^{\mu\nu})^2+\frac{1}{2}m^{2}_{\rm X} (X^{\mu})^2+\frac{1}{2}\chi F^{\mu\nu}X_{\mu\nu}.
\end{eqnarray}
We note that, at Belle II the energy is $E\ll m_{Z}$ and we can safely ignore the mixing with the $Z$ boson.
Therefore, as usual, $F^{\mu\nu}$ is the electromagnetic field strength. $X^{\mu}$ is the massive ($m_{X}$) dark photon with associated field strength $X^{\mu\nu}$. The interaction with the SM is provided by the kinetic mixing parameter $\chi$. 

\begin{figure}[t]
	\centering
	\includegraphics[width=0.49\linewidth]{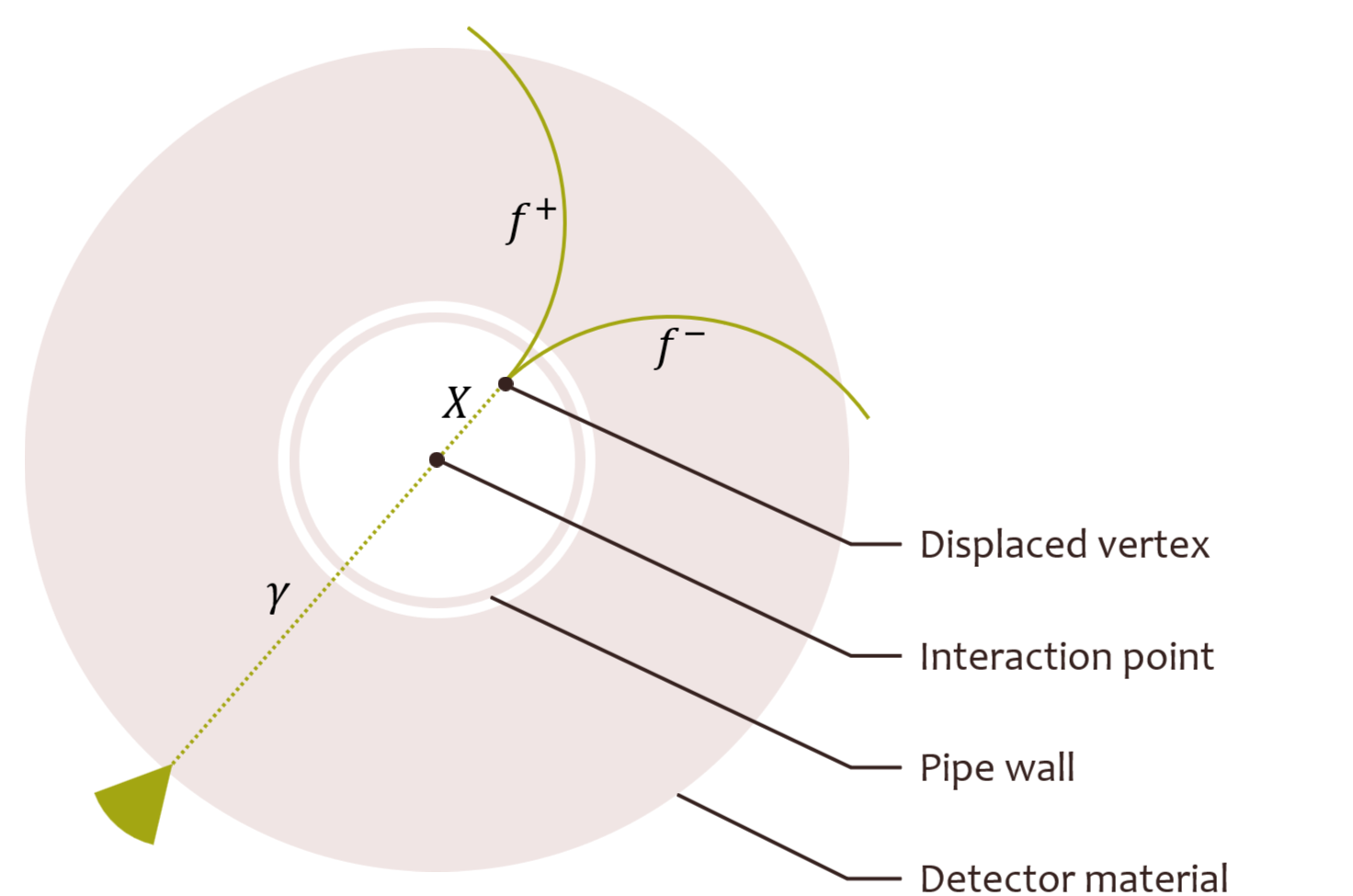}
	\includegraphics[width=0.49\linewidth]{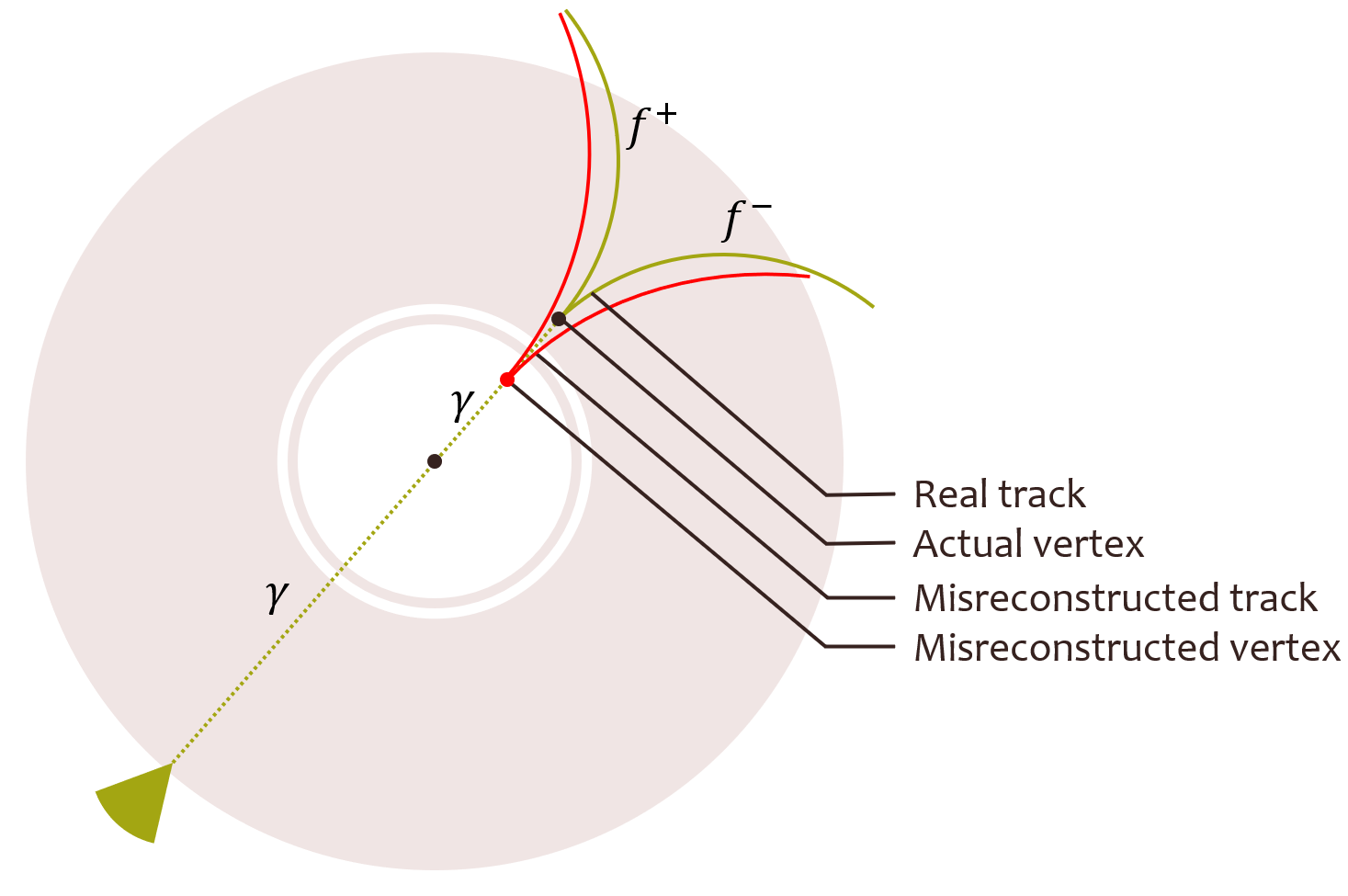}
	\includegraphics[width=0.49\linewidth]{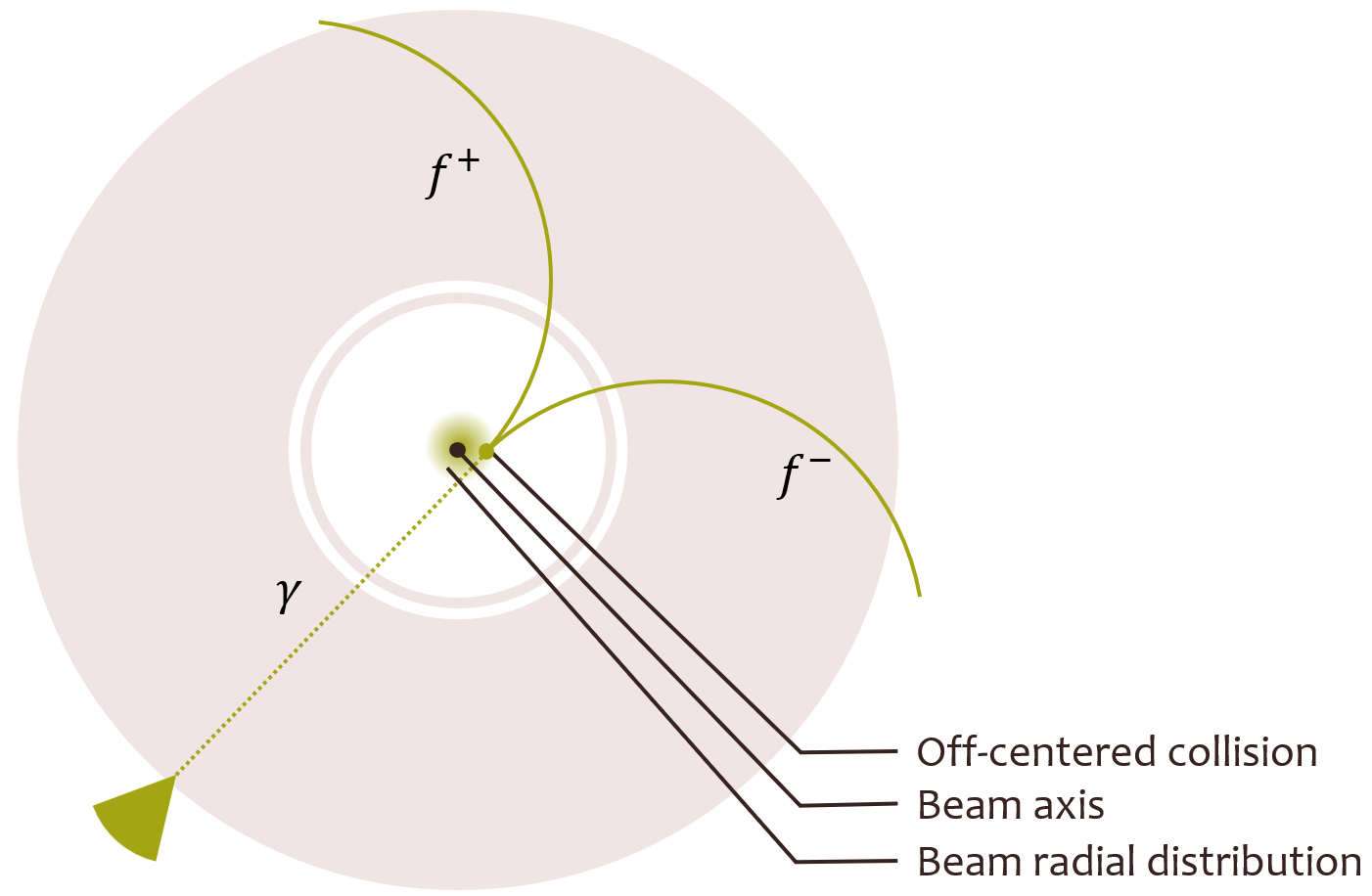}
	\includegraphics[width=0.49\linewidth]{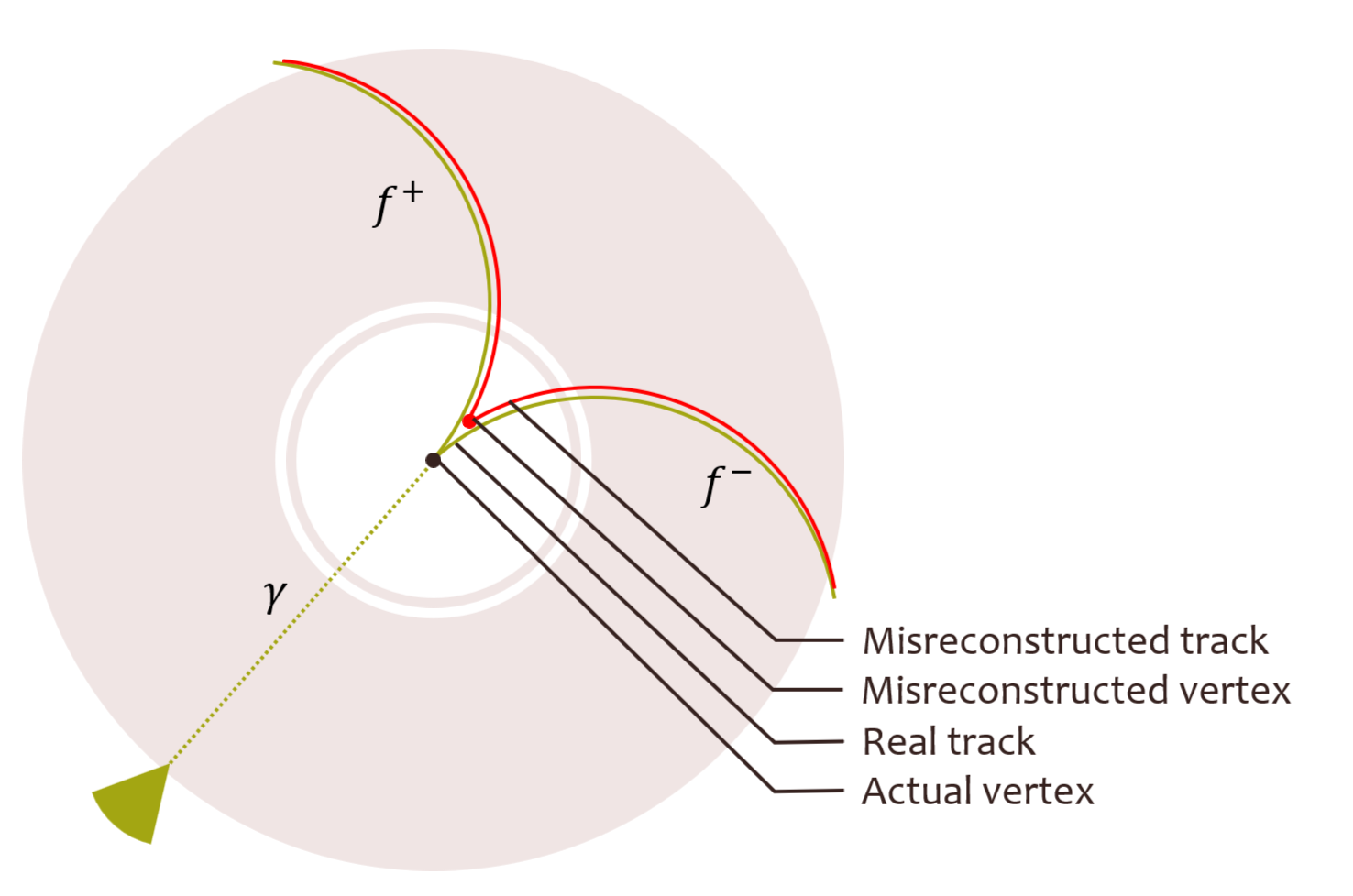}
	\caption{{\bf Top left:} Schematic view of displaced vertices in the Belle-II detector (the shown plane is perpendicular to the beam axis). {\bf Top right:} Schematic view of a mis-reconstructed photon conversion event. {\bf Bottom left:} Off-centered prompt events are a background for searches within $ R < 0.2$ cm. {\bf Bottom right:} Mis-reconstructed prompt events are also a source of background near the beam axis. $ f = e,\mu,\pi,K. $.}
	\label{fig:dpscheme}
\end{figure}

At Belle II, dark photons can be produced via the process $e^+ e^- \to X \gamma$ and can be detected via their decay $X \to f^+ f^-$ with $f = e,\mu,\pi,K$. For suitable mass and kinetic mixing, the dark photon can travel a macroscopic distance before its decay. The signal that we are looking for is a pair of oppositely charged particles originating from a displaced vertex and is balanced by a high energy photon. The energy and momentum of the two colliding $e^+ e^-$ beams must be equal to the energy and momentum of the three visible particles. A schematic of such displaced vertices is shown in the top left panel of \figref{fig:dpscheme}. The number of signal events obeys (cf., e.g.~\cite{Balkin:2021jdr} from whom we have adapted the formula)
\begin{align}
	S_{f\bar{f}} = \int L \rm dt \int \rm d\cos\theta^{\text{lab}} \dv{\sigma_{e^+ e^- \to \gamma X}}{\cos\theta^{\text{lab}}} \qty(e^{-\frac{\rmin}{\lambda_X \cos\theta^{\text{lab}}}}-e^{-\frac{\rmax}{\lambda_X \cos\theta^{\text{lab}}}}) \text{Br}(X\to f\bar{f}) \mathcal A. \label{eq:dpsignal}
\end{align}
As usual the branching ratio of interest is defined as
\begin{align}
    \text{Br}(X\to f\bar{f}) = \frac{\Gamma(X\to f\bar{f})}{\Gamma_X},
\end{align}
where $\Gamma_X$ is the total decay width of dark photons,
\begin{align}
    \Gamma_X = \sum_{2m_l \leq m_X} \Gamma(X \to l^+ l^-) + \Gamma(X \to \text{hadrons}).
\end{align}
Here, $\Gamma(X \to l^+ l^-)$ and $\Gamma(X \to \text{hadrons})$ are the leptonic and hadronic partial decay width of dark photons. The latter can be computed (see~\cite{Andreas:2012mt}) using the R-ratio as measured in \cite{ParticleDataGroup:2022pth} via
\begin{align}
    \Gamma(X\to \text{hadrons}) = \Gamma(X \to \mu^+ \mu^-) R (\sqrt{s} = m_X).
\end{align}
For specific decay channels such as $X \to h^+h^-$, where $h = \pi, K$, we make use of the form factors $F_h(q^2)$ as can be found in~\cite{Bruch:2004py}
\begin{align}
	\Gamma(X\to h^+ h^-) = \frac1{12} \alpha \chi^2 m_X^2 \abs{F_h(q^2)}^2 \qty(1 - \frac{4 m_h^2}{m_X^2})^{3/2}. \label{eq:Xtohh}
\end{align}

\bigskip

\subsection{Relevant search regions and selection criteria}
\label{sec:selection-criteria}
A thorough discussion of the sensitivity of Belle II to this kind of signal as well as possible selection criteria can be found in~\cite{Ferber:2022ewf}. Here, we briefly summarize their main points and highlight some differences to our analysis. 

\subsubsection*{Search regions}

{\bf $\mathbf{R<0.2}$~cm:}
As pointed out in Ref.~\cite{Ferber:2022ewf}, dark photon decays within $R = \sqrt{x^2 + y^2} < 0.2$ cm of the interaction point suffer from a large prompt background, i.e. direct production of electron-positron pairs at the interaction point. Therefore, we, too, do not consider displaced vertex searches in this region. 

\bigskip

{\bf $\mathbf{0.2}$~cm $\mathbf{< R < 0.9}$~cm: } This is far enough from the collision point that the prompt SM background is expected to be small. In addition, this region is in vacuum and there is no background from conversion of photons $\gamma \to f^+ f^-$ in material. Off-center collisions, as depicted in the bottom left panel of \figref{fig:dpscheme}, are expected to be negligible. However, this region is affected by backgrounds from the $\gamma \to f^+ f^-$ conversion in material due to potential mis-reconstructions of tracks that start in the material of the vertex detector (cf. red lines in the bottom right panel of \figref{fig:dpscheme}). 
Taking this into account we find that nevertheless this will be the most sensitive region for dark photon searches (this is in line with~\cite{Ferber:2022ewf}).

\bigskip

{\bf $\mathbf{0.9}$ cm $\mathbf{< R < 60}$ cm:} In this outer region, the $\gamma \to f^+ f^-$ photon conversion background, as depicted by the green lines in the right hand part of \figref{fig:dpscheme},  is irreducible.  As we will see in Sec.~\ref{sec:background} (see in particular \figref{fig:background}), and unlike the expectation in \cite{Ferber:2022ewf}, the background for  $X \to e^{+}e^{-}$ is likely to be relevant and even $X \to \mu^{+}\mu^{-}$ is not fully negligible. Although we do not compute the background $X \to h^+ h^-$, for hadronic decays we expect this background to also be noticeable. It will turn out that these irreducible backgrounds are so large that this region will not yield useful sensitivity.

\bigskip

{\bf $\mathbf{R > 60}$ cm:} For this region, we do not have enough information for displaced vertices searches.

\subsubsection*{Selection criteria}

Similar to~\cite{Ferber:2022ewf}, we employ kinematic cuts that take into account the geometry of the Belle II detector. We also demand that the invariant mass of $f^+ f^-$ pair $m_{ff} > 0.03$ GeV. For simplicity, we do not impose cuts on $p(e^+), p(e^-), p_T (\mu^+), p_T(\mu^-)$ and the photon energy $E_{\rm{lab}}$, as they either have an insignificant effect or are redundant. We also do not exclude the mass region around the $K^0_S$ mass, since our background analysis is done only for leptons. 

\begin{figure}[t]
	\centering
	\includegraphics[width=0.7\linewidth]{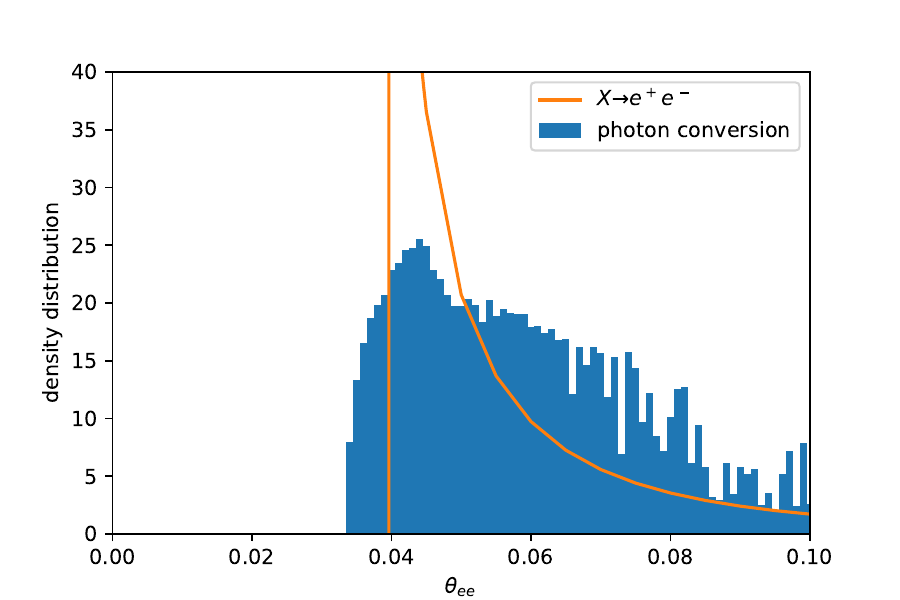}
	\caption{The $e^+ e^-$ opening angle density distribution of the dark photon decay signal at $m_X = 105$ MeV (orange) and the photon conversion background with 90 MeV $\leq m_{e^+ e^-} \leq$ 120 MeV (blue). Both distributions are normalized to 1. In both cases, $E_{e^+} + E_{e^-} = 5.3$~GeV. }
	\label{fig:openangle-CMangle}
\end{figure}

We also find that a lower cut on the opening angle $\theta_{ee}$ hurts sensitivity. Figure~\ref{fig:openangle-CMangle} shows both signal (orange line) and background (blue histograms). Except in the co-linear regime, $\theta_{ee}$ is kinematically bounded from below and concentrated at $\theta^{\text{peak}}_{ee}$ defined by
\begin{align}
	\sin \theta^{\text{peak}}_{ee} = \frac{2\sqrt{(\nu^2-m_{ee}^2)(m_{ee}^2-4m_e^2)}}{\nu^2-4m_e^2}, \label{eq:thetaff}
\end{align}
where, $\nu$ is the total energy of the $e^+ e^-$ pair in the lab frame. Assuming a uniform angular distribution in the center-of-mass frame of the $e^+ e^-$ pair, $\theta_{ee}$ peaks at $\theta^{\text{peak}}_{ee}$ in the lab frame. These statement are independent of the underlying dynamics, and therefore are also true for $e^+ e^-$ pair produced from photon conversion with the same invariant mass and total energy. Therefore, a lower cut on $\theta_{ee}$ above $\theta^{\text{peak}}_{ee}$ cuts most of the signal, while a lower cut below $\theta^{\text{peak}}_{ee}$ has little effect on the conversion background. However, the finite resolution of $m_{ff}$ suggests that the $\theta_{ee}$ distribution of the conversion background has a thicker tail at high $\theta_{ee}$, and therefore an upper-bound on $\theta_{ee}$ can improve the sensitivity.

\section{Backgrounds}
\label{sec:background}

As already indicated we mainly focus on the region $ R > 0.2$ cm. There, the relevant  background for our search are photons from $e^+ e^- \to \gamma \gamma$ where one photon converts to $f^+ f^-$ by interacting with matter via $\gamma N \to f^+ f^- N$, as identified in \cite{Ferber:2022ewf} and shown in the top right panel of \figref{fig:dpscheme}. 
In this process, $N$ denotes an atomic nucleus in the detector material. Since the nuclear recoil is not observable, we see the same final product as with the dark photon decay signal: a pair of $f^+ f^-$ recoiling against a high energy $\gamma$.
In the region $R> 0.9$~cm this background is irreducible (green lines).
But, we also note that even the region $0.2\,{\rm cm}<R< 0.9$~cm, which is in vacuum, is affected by photon conversion. The reason is that the reconstruction algorithm can mistakenly ascribe an $f^+ f^-$ pair originating from a conversion in matter to a region closer to the beam pipe as shown by the red lines in the top right panel of \figref{fig:dpscheme}. 

In addition, there are two types of prompt background shown in the lower panels of \figref{fig:dpscheme}. Due to the finite beam size, collisions may happen off-center, thereby looking like a displaced vertex (bottom left panel). Further, prompt production of electron-positron pairs can affect the displaced vertex search due to mis-reconstruction. Below we argue that both types of background are likely to be manageable.

Let us now turn to an actual calculation of the background rate and its properties. Here, we shall sketch the main points. Additional details are given in appendices~\ref{app:conversion} and~\ref{app:simulation}. 

\subsection{Simulation of the conversion background}
\label{sec:background-simulation}
The main point of this paper is a relatively careful treatment of the photon conversion background for the leptonic decay of dark photons, i.e. $f = e,\mu$. The case for hadronic final states shall be left for future studies. 

\begin{figure}[t]
	\centering
    \includegraphics[width=\linewidth]{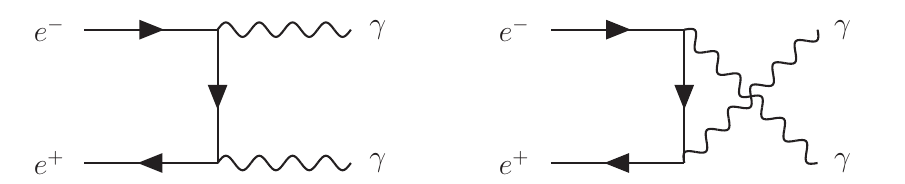}
	\caption{The main photon production processes from the SM.}
	\label{fig:photondiagrams}
\end{figure}

The first step in the calculation is the production of photons from SM processes. This proceeds via the diagrams shown in \figref{fig:photondiagrams}.

\begin{figure}[t]
	\centering
    \includegraphics[width=\linewidth]{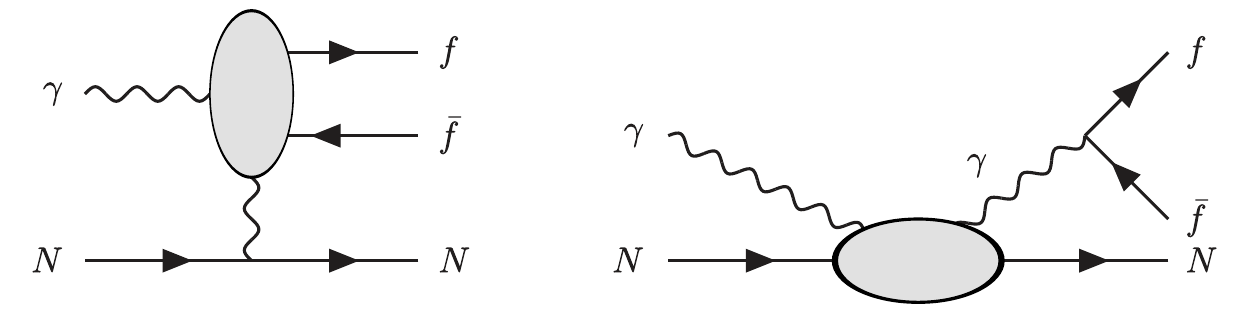}
	\caption{The diagrams involved in photon conversion. Left: Bethe-Heitler (BH). Right: timelike Compton scattering (TCS).}
	\label{diag:photon-conversion}
\end{figure}

These photons are then converted into lepton pairs via the processes shown in \figref{diag:photon-conversion}. A detailed calculation of the relevant cross section is given in Appendix~\ref{app:conversion}. Schematically, the cross section for conversion has the following schematic form,
\begin{align}
	d\sigma = d\sigma_{TCS} + d\sigma_{BH},
\end{align}
where the diagrams for timelike Compton scattering, $d\sigma_{TCS}$, and for Bethe-Heitler pair production, $d\sigma_{BH}$, are shown in \figref{diag:photon-conversion}. The interference term between these two contributions vanishes upon performing the relevant phase space integration.

\bigskip

To simulate the background, we use a model of the Belle II detector that consists of layers of concentric cylindrical shells. Each shell is made of a chemical element that represents each layer of the detector. The width of each shell is chosen to match the material budget presented in~\cite{Belle-II:2010dht}.
The detailed information we use is summarized in table~\ref{tab:layer-info} in Appendix~\ref{app:simulation}.

The number of background conversion events is
\begin{align}
	B^{\rm{conversion}}_{f\bar f} = \int L dt \int_{\text{fid}} d\cos\theta^{\text{lab}} \dv{\sigma_{e^+ e^- \to \gamma \gamma}}{\cos\theta^{\text{lab}}} p_{\gamma \to f \bar f}.
\end{align}
Here, $p_{\gamma \to f \bar f}$ is the probability for photon conversion and, if applicable, \linebreak mis-reconstruction. Summing the probabilities over the shells we have,
\begin{align}
    p_{\gamma \to f \bar f} = 2\sum_i p^{\text{bg}}_i\, p^i_{\gamma \to f \bar f}.
\end{align}
The factor of 2 takes into account that 2 photons are produced. $p^{\text{bg}}_i $ accounts for mis-reconstruction. For a photon conversion happens at layer $i$, a distance $r_i$ from the beam axis,
\begin{equation}
	p^{\text{bg}}_i = \bigg\{ \begin{array}{ll}
	1 &\text{for conversions in the search region},\\
    0 &\text{for conversions at smaller radius than the search region} \\
    \frac{e^{R_{\mathrm{max}}/\lambda} - e^{R_{\mathrm{min}}/\lambda}}{e^{r_i/\lambda}-1}\quad &\text{for conversions at larger radius than the search region\,.}
    \end{array}
    \label{eq:cuts}
\end{equation}
Here, $R_{\mathrm{min}}$ and $R_{\mathrm{max}}$ are the inner and outer limit of the search region. The assignment of the mis-reconstruction probability is taken to be asymmetric due to the tendency of the reconstruction algorithms to pull the vertex toward the center~\cite{Ferber:2022aaa}. Further, we assume an exponentially decaying profile with length $\lambda$ for inward mis-reconstruction.
In the limit $\lambda \to \infty$, we obtain a flat mis-reconstruction probability. Note that, if a photon conversion happens in the search region, we always include it as a background, despite mis-reconstruction. This causes some double counting but it is conservative.

$p^i_{\gamma \to f \bar f}$ is the probability that a photon converts in layer $i$
\begin{align} \label{eq:conversion-probability}
    p^i_{\gamma \to f \bar f} = \qty(\prod_{j=1}^{i-1} p^\gamma_j) (1 - p^\gamma_i) \,\frac1{\sigma_i} \int_\text{fid} d\sigma_{\gamma N_i \to f \bar f N_i}.
\end{align}
The calculation of the conversion cross section $\sigma_{\gamma N_i \to f \bar f N_i}$ is presented in appendix \ref{app:conversion}. $\sigma_i$ is the total photon-nucleus cross section and can be found in, for example, \cite{xcom:2010}. $p^\gamma_i$ is the probability that a photon survives passing through layer $i$ with thickness $z_i$ and number density $n_i$
\begin{align}
    p^\gamma_i = \exp(-n_i \sigma_i \frac{z_i}{\sin\theta^\text{lab}}).
\end{align}
The layer information, based on \cite{Belle-II:2010dht,Friedl:2013gta}, is shown in table~\ref{tab:layer-info} of Appendix~\ref{app:simulation}. 

To reduce background, we also consider the geometry of the Belle II detector and impose the following kinematical cuts that a dark photon signal automatically satisfies, based on \cite{Belle-II:2018jsg},
\begin{align}
	&\angle \qty(\vec q_\gamma, \vec p_{f^+} + \vec p_{f^-}) < 0.01 \text{ rad}\q
	\abs{\vec q_\gamma - \qty(\vec p_{f^+} + \vec p_{f^-})} < 0.1 \GeV,\nn
	&\abs{E_{f^+} + E_{f^-} - \nu} < 0.1 \GeV\q
	\abs{m_{\gamma'f^+ f^-} - \sqrt{s}} < 0.03 \GeV, \label{eq:em-conserve}
\end{align}
where $ l_+ = (E_{f^+}, \vec p_{f^+}) $ and $ l_- = (E_{f^-}, \vec p_{f^-}) $ are the 4-momenta of $ f^+ $ and $ f^- $, $ q = (\nu,\vec q_\gamma) $ and $ q' = (E'_\gamma,\vec q_\gamma') $ are the 4-momenta of the converted and the recoiled photons, $ m_{\gamma'f^+ f^-}^2 = \qty(q'+l_+ + l_-)^2 $, and $ \sqrt{s} = 10.58 \GeV $ for Belle II.

\subsection{Improving selection criteria}
\label{sec:background-reduction}
Since photon conversion requires momentum transfer with a nucleus, a significant part of the background can be removed by requiring that the three final state particles fulfill the energy momentum conservation relation with the initial beam up to the resolution of the detector.

To further reduce the background, we search for dark photons in one mass bin at a time. This is because the dark photon signal has a very sharp $m_{f^+ f^-}$ peak, while the photon conversion background has a spectrum of $m_{f^+ f^-}$.

In principle, this could be further improved. As discussed in section \ref{sec:selection-criteria}, the opening angle distribution at a certain dark photon energy and mass is peaked around a certain value $\theta^{\text{peak}}_{ff}$. Therefore, a mass and energy dependent opening angle cut concentrating around this peak should improve the confidence level. However, optimization is needed to obtain the best result and we leave this to future work.

\subsection{Prompt Backgrounds}

{\bf Finite beam size (bottom left of \figref{fig:dpscheme}):}
The prompt process that may mask our signal is the prompt $e^+ e^- \to f^+ f^- \gamma$. According to \cite{Belle-II:2018jsg}, the vertical and horizontal beam size of SuperKEKB, the accelerator that hosts the Belle II experiment, is about 50 nm and 10 $\mu$m respectively. Therefore, we do not expect the smearing of the beam to result in a relevant prompt background for $R > 0.2$ cm. 

\noindent {\bf Mis-reconstructed central events (bottom right of \figref{fig:dpscheme}):} Another source of potential background is from the mis-reconstruction of prompt $e^+ e^- \to f^+ f^- \gamma$. However, due to the tendency of the Belle II's reconstruction algorithm to pull vertices toward the center~\cite{Ferber:2022aaa}, we expect this type of background to also be negligible.
We nevertheless note that, as can be seen in \figref{fig:background} this requires a suppression on the level of 4 or more orders of magnitude (as can be seen from the comparison in Fig.~\ref{fig:RegionPlot-vary}) and therefore careful experimental validation.

\section{Results}
\label{sec:results}
In this section, we now present the results of the background simulation and the estimate for the dark photon parameter space coverage of Belle II. We consider  the different detector regions and various values for the mis-reconstruction length $\lambda$ and integrated luminosity.

Due to the large background we do no consider dark photons with masses below 30 MeV, following \cite{Ferber:2022ewf}. 

As will become clear, due to the overwhelming background, searches for displaced vertices in the region $R > 0.9$ cm do not visibly contribute to constraining the parameters of the dark photon model and are not considered.

\subsection{Background simulation}
\label{sec:background-result}
\begin{figure}[t]
	\centering
	\begin{subfigure}{0.48\linewidth}
		\centering
		\includegraphics[width=\linewidth]{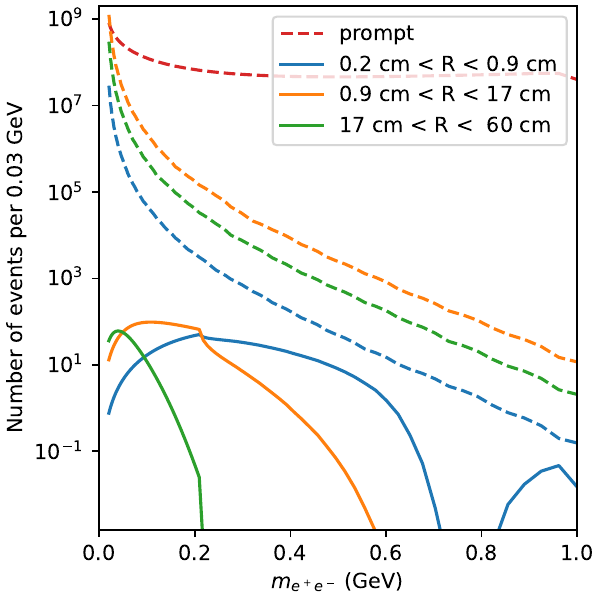}
	\end{subfigure}
	\begin{subfigure}{0.48\linewidth}
		\centering
		\includegraphics[width=\linewidth]{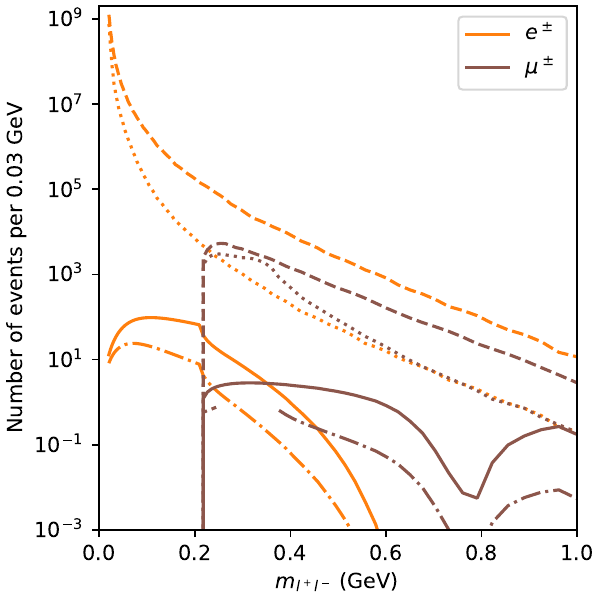}
	\end{subfigure}
	\caption{Expected number of signal with (dash-dotted) and without (solid) an opening angle cut, and background events with (dotted) and without (dashed) an opening angle cut at Belle II with integrated luminosity of $ 50 $ $ ab^{-1} $. The opening angle cut refers to the cut $\theta_{ll} \leq \theta^{\text{peak}}_{ll} + 5\times 10^{-3}$ with $\theta^{\text{peak}}_{ll}$ as defined in \eqref{eq:thetaff}. \textbf{Left:} for $X \to e^+ e^-$ at different regions with $\chi = 2\times 10^{-5}$. \textbf{Right:} for $X \to e^+ e^-$ (with $\chi = 2\times10^{-5}$) and $X \to \mu^+ \mu^-$ (with $\chi = 5\times 10^{-6}$) in the region $0.9$ cm $\leq R \leq 17$ cm.
 }
	\label{fig:background}
\end{figure}

The number of events for each type of background, with and without the opening angle cut is shown in \figref{fig:background}. As can be seen, displaced vertex searches for $R > 0.9$~cm  have large numbers of background events. It turns out that the background is always larger than the signal for all interesting values of the parameter space.

Let us briefly take a closer look at the sub-regions considered in~\cite{Ferber:2022ewf}.

In the region $0.9$~cm~$< R < 17$~cm, the authors of \cite{Ferber:2022ewf} expect a prohibitively large background for $X \to e^+ e^-$ and a negligible background for $X \to \mu^+ \mu^-$. However, we find that this region is not viable for both final states due to the non-negligible conversion background.

In the region $17$~cm~$\leq R < 60$~cm,~\cite{Ferber:2022ewf} assumed that the background for $X \to e^+ e^-$ is small. However, \figref{fig:background} shows that for $m_X < 0.7$~GeV, which is the region where $X\to e^+ e^-$ is relevant, we expect $> \mathcal O(10^2)$ background events per mass bin. This gets significantly worse when we go to lower masses. We remind the reader that the cut on the opening angle imposed by the paper does not improve the significance of the signal. Similarly, \cite{Ferber:2022ewf} assumes the background for $X \to \mu^+\mu^-$ to be negligible, but we find that the channel can have between $\mathcal O (10)$ background events at $m_X = 1$ GeV to $\mathcal O (10^4)$ background events at $m_X = 300$ MeV, which is not quite negligible. Although we did not simulate the background for $X \to h^+ h^-$, we expect this background to also be sizeable. These backgrounds render searches in $R > 0.9$~cm not viable. Therefore, in the following Sec.~\ref{sec:BelleII-coverage} we focus on searches in the beam. 
Here, too, it turns out that the background is always larger than the signal for all interesting values of the parameter space.

As already argued in Sec.~\ref{sec:selection-criteria} a cut on the opening angle centered around the peak opening angle, defined in \eqref{eq:thetaff}, has non-trivial effects on both signal and background and requires optimization. This can also be seen from the right hand side panel of \figref{fig:background}.
Therefore, in the following we do not consider a cut on opening angle.

\subsection{Coverage of Belle II assuming different degree of misreconstruction}
\label{sec:BelleII-coverage}
\begin{figure}[t]
\centering
		\includegraphics[width=0.48\linewidth]{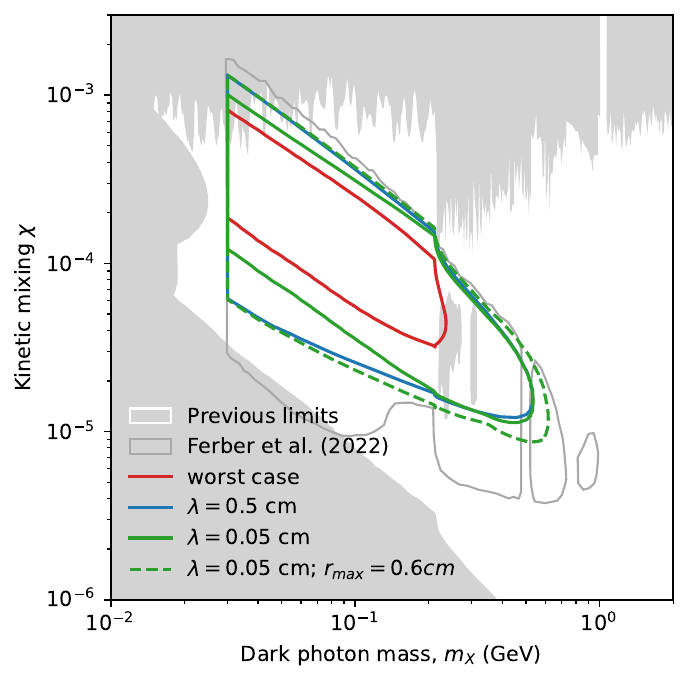}
		\includegraphics[width=0.48\linewidth]{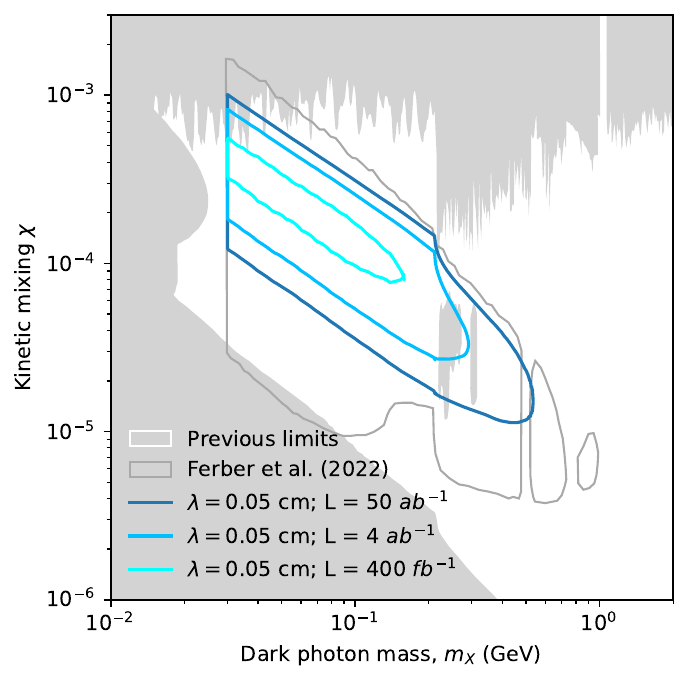}
\caption{90\% C.L. regions for a long-lived dark photon search using displaced vertices in the beam pipe of Belle II. We assume an exponential profile with length $\lambda$ for the mis-reconstruction of the conversion background and a prompt background suppression of $10^{-6}$ unless stated otherwise. Previous limits are shown in gray and the sensitivities obtained in~\cite{Ferber:2022ewf} are indicated by gray lines. {\bf Left:} Various mis-reconstruction decay lengths $\lambda$ are shown. The integrated luminosity is 50 ab$ ^{-1} $. The worst case (red) refers $\lambda \to \infty$ and a mis-reconstruction probability of $10^{-4}$ for the prompt background. The dashed green line shows a situation where the search region is shrunk to [0.2~cm,~0.6~cm]. {\bf Right:} Various integrated luminosities as indicated for $\lambda = 0.05\,{\rm cm}$.}  
	\label{fig:RegionPlot-vary}
\end{figure}

In this section, we present the parameter space coverage of Belle II, taking into account the simulated background. As discussed in Sec.~\ref{sec:background-result}, the background in the region $ R  > 0.9$ cm is not negligible. Because of this, including searches in this region does not give noticeable improvement to the final result. We, therefore, do not include them in \figref{fig:RegionPlot-vary}. 

In the region $0.2$ cm $\leq R \leq 0.9$ cm, the main background is mis-reconstructed photon conversion events. To parameterize this, we assume an exponentially decaying mis-reconstruction profile with a decay length $\lambda$, as described in \eqref{eq:cuts}. More details can be found in App.~\ref{app:simulation}. The prompt $e^+ e^- \to f^+ f^- \gamma$ can also be a potential source of background. To take this into account, we assume that a flat percentage of prompt background is mis-reconstructed into our search region. Unless stated otherwise, we take this percentage to be $10^{-6}$.

Figure \ref{fig:RegionPlot-vary} indicates the predicted coverage of Belle II for different \linebreak mis-reconstruction  lengths $\lambda$ and integrated luminosities $L$. As can be seen, an improvement of the reconstruction algorithm, as demonstrated by a smaller value of $\lambda$, can have significant impact on the performance of Belle II dark photon searches (left panel). Specifically, a mis-reconstruction decay length $\lambda = 0.05$ cm is enough to discern dark photon signals with the current integrated luminosity of Belle II (see right panel). For larger luminosities the coverage of so far untested parameter space increases significantly. Shrinking the search region to, for example, $0.2$~cm~$\leq R \leq 0.6$~cm improves the sensitivity as it reduces the background from mis-reconstructed conversion events, see the dashed green line in~\figref{fig:RegionPlot-vary}.

Our sensitive region is generally smaller than that of~\cite{Ferber:2022ewf}, but the sensitivity loss is not dramatic. We lose sensitivity in the region that corresponds to $R \geq 17$ cm in~\cite{Ferber:2022ewf}, i.e. the region below $2 m_\mu$ with $\chi < 10^{-4}$, because searching there is no longer viable due to conversion background. We have less sensitivity for $m_X > m_\pi$ because we did not take hadronic decays of dark photons into account. Also due to that, we include the mass region around $K^0_S$ since leptonic decays of $K^0_S$ are very suppressed.

\section{Conclusions}
\label{sec:conclusion}
In this work, we studied the prospects for probing kinetically mixed dark photons by searching for displaced vertices in the Belle II detector. We improved upon previous studies~\cite{Ferber:2022ewf} (see also~\cite{Bandyopadhyay:2022klg,Bandyopadhyay:2023lvo}) by a more detailed investigation of the backgrounds. 

The main signal for dark photons is from the process $e^+ e^- \to \gamma X$ with a subsequent delayed decay $X \to f^+ f^-$, resulting in a pair of displaced tracks and a high energy photon. 
Similar to previous works, we considered the region 0.2~cm~$ < R < $~60~cm.
In this region prompt backgrounds are expected to be relatively small\footnote{We note that this depends on a non-trivial feature of the reconstruction algorithm that makes it unlikely that prompt tracks are reconstructed as displaced~\cite{Ferber:2022aaa}. Since the prompt background is quite large we also briefly discussed the impact of not complete rejection.}. 
However, taking into account the Standard Model background from photon conversions into lepton pairs inside the detector material (calculated in the present work) renders most of this region insensitive. This leaves the 0.2~cm~$<R<$~0.9~cm vacuum region inside the beam pipe. Despite this region being in vacuum, the pair conversion in the detector material further outside leads to a background from mis-reconstruction. This background is in principle reducible.
In any case, already for modest rejection of this background, significant new areas of parameter space can be probed at Belle II.

\acknowledgments
We are deeply indebted to Torben Ferber for valuable discussions. JJ acknowledges support from the European Union via the Horizon 2020 research and innovation programme under the Marie Sklodowska-Curie grant agreement No 860881-HIDDeN. Part of this work was done in the context of AVP's Master thesis and AVP gratefully acknowledges support for this via the Deutschlandstipendium.

\appendix
\section{Computation of photon conversion cross section}\label{app:conversion}
The photon conversion process
\begin{align}
	\gamma(q) + N(p) \to f^-(l_-) + f^+(l_+) + N(p')
\end{align}
can be described by the 2 diagrams shown in \figref{diag:photon-conversion} for the case $ f = e,\mu $. Following \cite{20.500.12030_5261}, we define
\begin{align}
	q' &= l_- + l_+,\\
	s &= (p + q)^2 = M^2 + 2 M\nu,\\
	t &= (p-p')^2,\\
	m_{ff}^2 &= {q'}^2 = (l_++l_-)^2,
\end{align}
where $ p^2 = {p'}^2 = M^2 $ and $ \nu $ is the energy of the incoming photon in the lab frame. The amplitude $ \mM $ for the process is the sum of the BH amplitude $ \mM_{BH} $ and the TCS amplitude $ \mM_{TCS}. $

Using the nucleus form factor \cite{DeVries:1987atn,20.500.12030_5261}, the tree-level BH amplitude is given by
\begin{align}
	\mM_{BH} = \frac{ie}{t} \varepsilon_\mu(q) \bar N(p') F_1(t) \gamma_\nu N(p) T^{\mu\nu}_{\gamma^*\gamma \to f\bar f},
\end{align}
where, for the case of final leptons, the pair production tensor is
\begin{align}
	T^{\mu\nu}_{\gamma^*\gamma \to l^+ l^-} = e^2\bar u(l_-) \qty[\gamma^\mu \frac{(\slashed{l}_- - \slashed q + m)}{-2l_- q} \gamma^\nu + \gamma^\nu \frac{(\slashed{q} - \slashed l_+  + m)}{-2l_+ q} \gamma^\mu] v(l_+).
\end{align}
For the case of final hadrons, an explicit formula for $ T^{\mu\nu}_{\gamma\gamma \to f\bar f} $ is rather complicated and is left for future studies. The TCS contribution is given by \cite{20.500.12030_5261}
\begin{align}
	\mM_{TCS} = \frac{ie^3}{m_{ff}^2} J_\nu (q') \varepsilon_\mu (q) \bar N(p') \qty(\frac1\alpha T^{\mu\nu}_{fTCS}) N(p),
\end{align}
with $ T^{\mu\nu}_{fTCS} $ being the quasi-real-forward TCS amplitude (fTCS), and the final current $ J_\nu(q') $ in the case of a lepton final state is
\begin{align}
	J_\nu (q') = e \bar u(l_-) \gamma_\nu v(l_+),
\end{align}
and in the case of $ h = \pi,K $ final states is
\begin{align}
	J_\nu (q') = e q'_\nu F_h({q'}^{2}).
\end{align}
The form factor $F_h({q'}^{2})$ can be found in, for example, \cite{Bruch:2004py}. In the near-real, near-forward limit $ m_{ff}^2, \abs{t} \ll s $, one can extract the Lorentz structure of the tensor \cite{Tarrach:1975tu,20.500.12030_5261}
\begin{align}
	T^{\mu\nu}_{fTCS} \approx \qty(-g^{\mu\nu} + \frac{{q'}^{\mu} q^\nu}{q \cdot q'}) T(\tilde\nu, t, {q'}^{2}),
\end{align}
where $ \tilde \nu = q \cdot (p+p')/(2M) $ approaches the incoming photon energy $ \nu $ in the limit $ M \to \infty, $ and $ T(\tilde\nu, t, {q'}^{2}) $ is the spin-averaged near-forward TCS amplitude~\cite{Tarrach:1975tu,20.500.12030_5261},
\begin{align}
	T(\tilde\nu, t, {q'}^{2}) \approx f(\tilde\nu),
\end{align} 
with the spin-averaged forward real Compton amplitude $ f(\tilde\nu) $. Using the optical theorem, one can write \cite{Gryniuk:2015eza}
\begin{align}
	\text{Im}\, f(\nu) = \frac{\nu}{4\pi} \sigma_{\gamma N}(\nu).
\end{align}
Here, $ \sigma_{\gamma N}(\nu) $ is the unpolarized cross section of total photoabsorption of nucleus $ N $. The analytic and crossing properties of $ f $ permit us to write
\begin{align}
	\text{Re}\,f(\nu) = - \frac{Z^2 \alpha}{M} + \frac{\nu^2}{2\pi^2} \fint_0^\infty d\nu' \frac{\sigma_{\gamma N}(\nu')}{{\nu'}^2-\nu^2}, \label{Ref-TCS}
\end{align}
where $ Z $ is the atomic number, and the slashed integral means the principal value integration. The first term above is the elastic scattering contribution, while the remaining terms in $ f(\nu) $ is inelastic. In the energy range of interest to us, the inelastic part of $ f(\nu) $ is one to two orders of magnitude greater than the elastic part; therefore, considering only the elastic contribution is not possible. With this, we have the means to compute the TCS contribution if we have the photoabsorption cross section for all relevant elements. Fortunately, as shown in \cite{Hutt:1999pz,Ahrens:1985hxw}, the photoabsorption cross section per nucleon $ \sigma_{\gamma N}/A $, with $ A $ being the number of nucleons, is approximately the same for all nuclei for $ \nu \gtrsim m_\pi \approx 140 \MeV $. Thus, for $ \nu > 4 \GeV $, which is the relevant photon energy in our experiment, we can obtain the photoabsorption cross section for all elements using only that for protons found in the same papers.

Using the amplitudes above, we can compute the differential cross section. It consists of 3 terms
\begin{align}
	d\sigma = d\sigma_{TCS} + d\sigma_{BH} + d\sigma_{INF},
\end{align}
which corresponds to the TCS amplitude squared, the BH amplitude squared, and the interference between these 2 amplitudes. Our calculation is further simplified as the interference term should vanish after phase space integration. Following \cite{Pire:2009ev}, we can argue that a charge conjugation of the $ f \bar f $ pair flips the sign of the TCS amplitudes, because it contains a single charge factor, while keeping the BH amplitude intact, because it features a charge squared factor. Thus, the interference term changes sign under such charge conjugation, and vanishes if our phase space integration is symmetric with respect to exchanging $ f $ and $\bar f $. This is always the case for our calculations. We have also checked the vanishing of the integrated interference term numerically.

\begin{figure}[t]
	\centering
	\begin{subfigure}{0.46\linewidth}
		\centering
		\includegraphics[width=\linewidth]{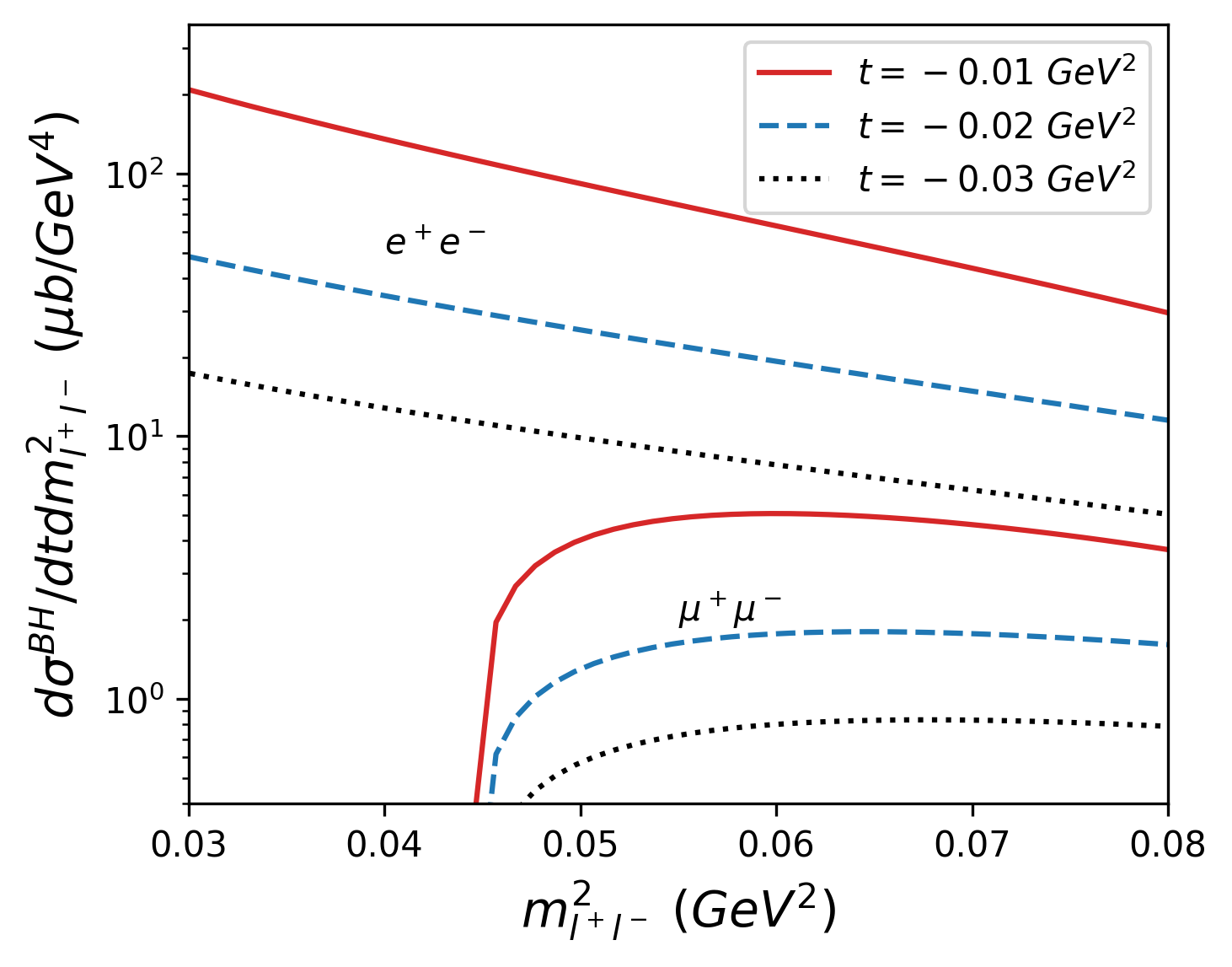}
		\label{fig:1804.03501.3}
	\end{subfigure}
 \hspace*{0.3cm}
	\begin{subfigure}{0.46\linewidth}
		\centering
		\includegraphics[width=\linewidth]{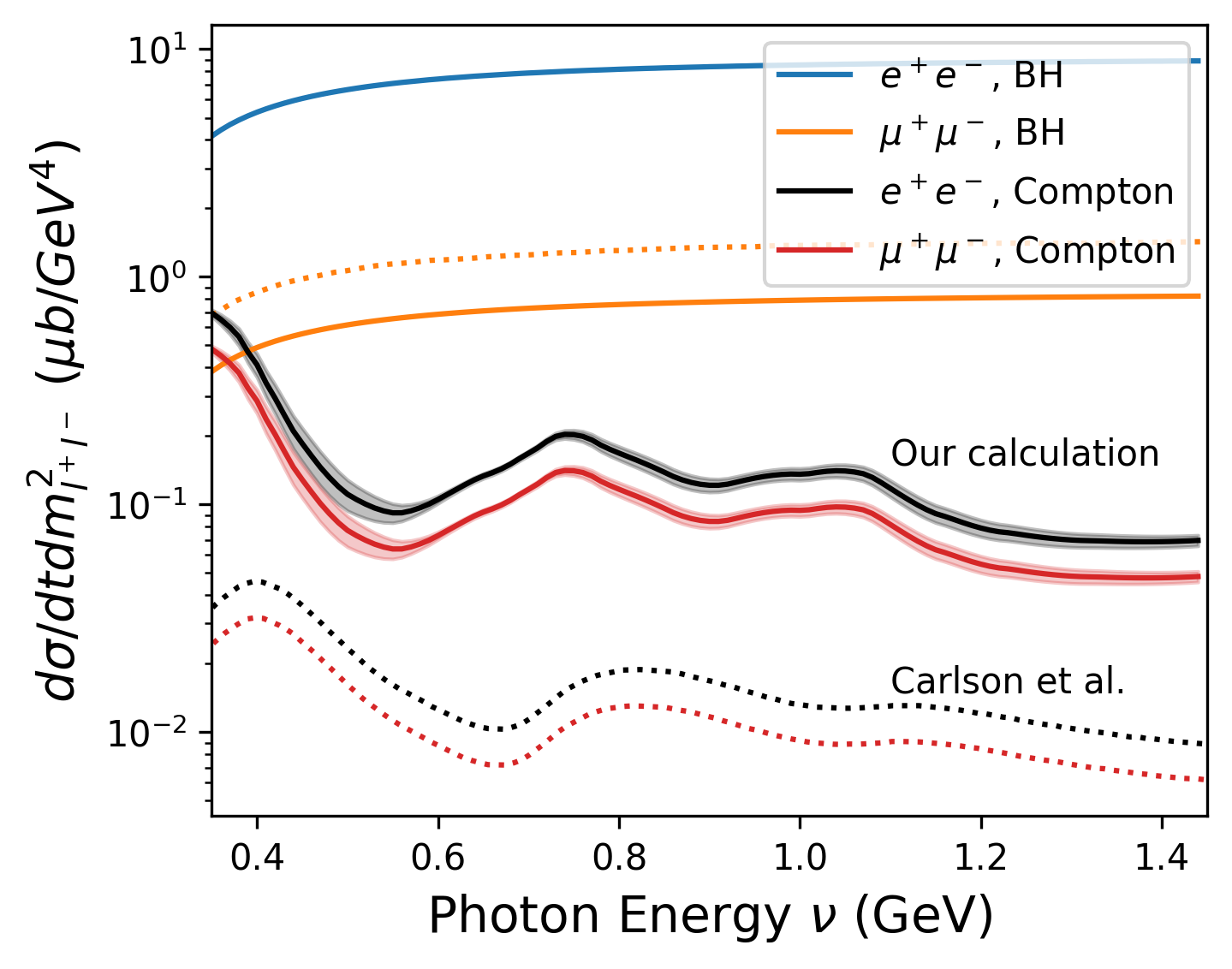}
		\label{fig:1804.03501.4}
	\end{subfigure}
	\caption{Different contributions to the differential cross section for $ \gamma d \to l^+ l^- d$ for different parameters. These figures are to be compared with \cite{Carlson:2018ksu}. In the left panel, the dependence of the differential cross section to the lepton pair invariant mass $ m_{l^+ l^-}^2 $ at photon energy $ \nu = 0.65 \GeV $ is plotted. This is to be compared with figure 3 in \cite{Carlson:2018ksu}. In the right panel, the dependence of the differential cross section to the photon energy $ \nu $ at $ t = - 0.03 \GeV^2 $ and $ m_{l^+ l^-}^2 = 0.06 \GeV^2. $ The black and red band are the computational uncertainties. This panel is to be compared with figure~4 in \cite{Carlson:2018ksu}, whose results are shown as dotted lines for comparison.}
	\label{fig:1804.03501}
\end{figure}

To check our calculation, we computed the differential cross section for photon conversion on deuterium $\gamma d \to l^+ l^- d$ and compared the result to~\cite{Carlson:2018ksu}. This is shown in \figref{fig:1804.03501}. We find a discrepancy between our calculation and that of~\cite{Carlson:2018ksu}. This can be seen in the right panel. 
The larger BH contribution is somewhat smaller in our calculation. The difference is nearly a factor of two when comparing the muon case (see the solid and dashed orange line in the right panel of \figref{fig:1804.03501}). Note, however, that comparing the point $ \nu=0.65 \GeV $ at $ t = - 0.03 \GeV^2 $ and $ m_{l^+ l^-}^2 = 0.06 \GeV^2$ between figures 3 and 4 of~\cite{Carlson:2018ksu} there also seems to be an internal difference by a factor of about two.  For the Compton contribution the relative discrepancy seems to be larger. However, it is a much smaller part of the total background. For the electron case and for the left panel, we find good agreement (on the level of 10\%) in the dominant contributions from BH.

\section{Simulation of photon conversion in Belle II}\label{app:simulation}

\begin{figure}[t]
	\centering
		\includegraphics[width=\linewidth]{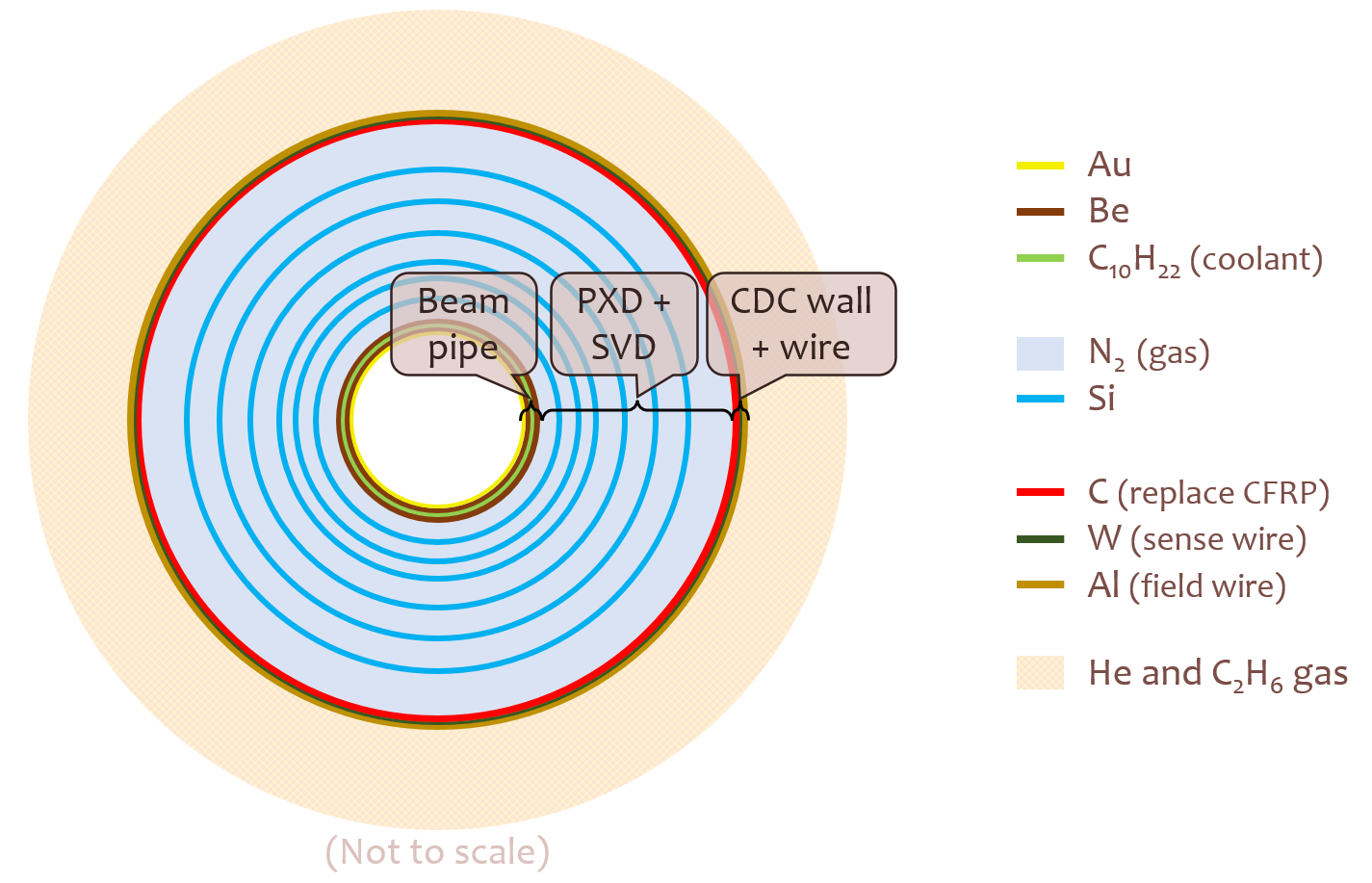}
	\caption{Schematic view of the Belle II detector in our simulation (not to scale).}
	\label{fig:schematic-BelleII}
\end{figure}

To simulate the photon conversion background, we use a model of the Belle II detector as shown schematically in \figref{fig:schematic-BelleII}. For simplicity, we consider the detector as concentric cylindrical shells of suitable materials and thickness. The shells are placed at the same radii as the detector components they represent, with the exception of the shells representing the sense wire and field wire which are both placed at a radius of 17 cm. Each shell is made of the dominant element, and has a thickness that yields the correct material budget, if available, or cross sectional area, if the former is unavailable. The detailed values are collected in table~\ref{tab:layer-info} for convenience.

\begin{table}[t]
	\centering
	\begin{tabular}{|c|c|c|c|c|c|}
		\hline
		$ r_i $ ($ cm $)& $ N_i $ & $ z_i $ ($ cm $) & $ X/X_0 $ & $ n_i $ ($ cm^{-3} $) &Note \\
		\hline
		1 & Au & $ 10^{-3} $ & \multirow{4}{*}{1\%} & $ 5.56371\e{22} $ &\multirow{4}{5cm}{\centering Beam pipe section. C represents liquid coolant.}\\
		1 & Be & $ 0.06 $ & & $ 1.22728\e{23} $ &\\
		1.06 & C & $ 0.1 $ & & $ 1.12497\e{23} $ &\\
		1.16 & Be & $ 0.04 $ & &  &\\
		\hline
		1.2 & N & 0.2 & & $ 0.054\e{21} $ &\multirow{13}{5cm}{\centering N represents air. Si represents Pixel Vertex Detector (PXD) or Silicon Vertex Detector (SVD)}\\
		1.4 & Si & $ 18.7\times 10^{-3} $ & 0.2\% & $ 4.95781\e{22} $ &\\
		1.4 & N & 0.8 & & &\\
		2.2 & Si & $ 18.7\times 10^{-3} $ & 0.2\% & &\\
		2.2 & N & 1.6 & & &\\
		3.8 & Si & $ 0.056 $ & 0.6\% & &\\
		3.8 & N & 4.2 && &\\
		8 & Si & $ 0.056 $ & 0.6\% & &\\
		8 & N & 3.5 && &\\
		11.5 & Si & $ 0.056 $ & 0.6\% & &\\
		11.5 & N & 2.5 & & &\\
		14 & Si & $ 0.056 $ & 0.6\% & &\\
		14 & N & 2 & &&\\
		\hline
		16 & C & 0.05 & & $ 1.12497\e{23} $ &\multirow{3}{5cm}{\centering CDC inner wall and wires.}\\
		17 & W & $ 9.64\times10^{-4} $ && $ 6.3222\e{22} $ &\\
		17 & Al & $ 0.0524 $& & $ 6.02627\e{22} $ &\\
		\hline
		17 & C & 44& & $ 2.26854\e{19} $ &\multirow{2}{5cm}{\centering Gases inside the CDC.}\\
		17 & He & 44& & $ 6.80728\e{19} $ &\\
		\hline
	\end{tabular}
	\caption{Information on the material layers of the detector. The layers are listed from the innermost to outermost and $r_i$ gives the starting point of each layer when going towards the outside. $N_i$ is the dominant element and $z_i$ is the thickness of the layer. Except for C, which has multiple relevant states, other elements have only 1 value for the number density $n_{i}$; thus, the density is shown only once for clarity. The thickness, and therefore the material budget ($ X/X_0 $) in terms of the radiation length, of each layer is chosen to match those of \cite{Belle-II:2010dht}.}
	\label{tab:layer-info}
\end{table}

\bibliographystyle{utphys}

\bibliography{ref} 

\end{document}